\documentclass{article}

\usepackage{arxiv}

\usepackage[utf8]{inputenc} 
\usepackage[T1]{fontenc}    
\usepackage{hyperref}       
\hypersetup{
    colorlinks=true,
    linkcolor=cyan,
    filecolor=magenta,      
    urlcolor=blue,
}
\usepackage{url}            
\usepackage{booktabs}       
\usepackage{amsfonts}       
\usepackage{nicefrac}       
\usepackage{microtype}      
\usepackage{lipsum}
\usepackage{graphicx}
\usepackage{amsmath}
\usepackage{multirow}
\usepackage{cite}

\title{Data Augmentation at the LHC through Analysis-specific Fast Simulation with Deep Learning}

\begin{document}

\author{
  C.~Chen \\
  Peking University \\
  Haidian, China, 100871
  \And
  O.~Cerri, T.~Q.~Nguyen, J.R.~Vlimant\\
  California Institute of Technology \\
  Pasadena, CA 91125, USA 
  \And
  M.~Pierini\\
  European Organization for Nuclear Research (CERN) \\
  CH-1211 Geneva 23, Switzerland
}

\maketitle

\begin{abstract}
  We present a fast simulation application based on a Deep Neural Network, designed to create large analysis-specific datasets. Taking as an example the generation of $W$+jet events produced in $\sqrt{s}=$~13~TeV proton-proton collisions, we train a neural network to model
  detector resolution effects as a transfer function acting on an analysis-specific set of relevant features, computed at generation level, i.e., in absence of detector effects. Based on this model, we propose a novel fast-simulation workflow that starts from a large amount of generator-level events to deliver large analysis-specific samples. The adoption of this approach would result in about an order-of-magnitude reduction in computing and storage requirements for the collision simulation workflow. This strategy could help the high energy physics community to face the computing challenges of the future High-Luminosity LHC.
\end{abstract}



\section{Introduction}

At the CERN Large Hadron Collider (LHC), high-energy proton-proton collisions are studied to consolidate our understanding of physics at the energy frontier and possibly to search for new phenomena. While these studies are typically conducted according to a {\it data driven} methodology, synthetic data from simulated proton-proton collisions are a key ingredient to a robust analysis development.  
Particle physicists rely extensively on an accurate simulation of the physics processes under study, including a detailed description of the response of their detector to a given set of incoming particles. These large sets of synthetic data are typically generated with experiment-specific simulation software, based on the {\tt GEANT4}~\cite{Agostinelli:2002hh} library. Through Monte Carlo techniques, {\tt GEANT4} provides the state of the art in terms of simulation accuracy. The first two runs of the LHC highlighted the remarkable agreement between data and simulation, with discrepancies observed at the level of a few percent. On the other hand, running {\tt GEANT4} is demanding in terms of resources. As a consequence of this, 
delivering synthetic data at the pace at which the LHC delivers real data is one of the most challenging tasks for the computing infrastructures of the LHC experiments. It is then more and more common for LHC physics analyses to be affected by large systematic uncertainties due to the limited amount of simulated data. This is particularly true for precise measurements of Standard Model processes for which large datasets are already available today.  In the future, with the high-luminosity LHC upgrade, this will become a serious problem for most of the LHC data analysis~\cite{Alves:2017she}. Our community is called to reduce the computing resources needed for central simulation workflows by at least one order of magnitude, not to jeopardize the accuracy gain expected when operating the LHC at a high luminosity.

\begin{figure}[t!]
    \centering
    \includegraphics[width=0.95\textwidth]{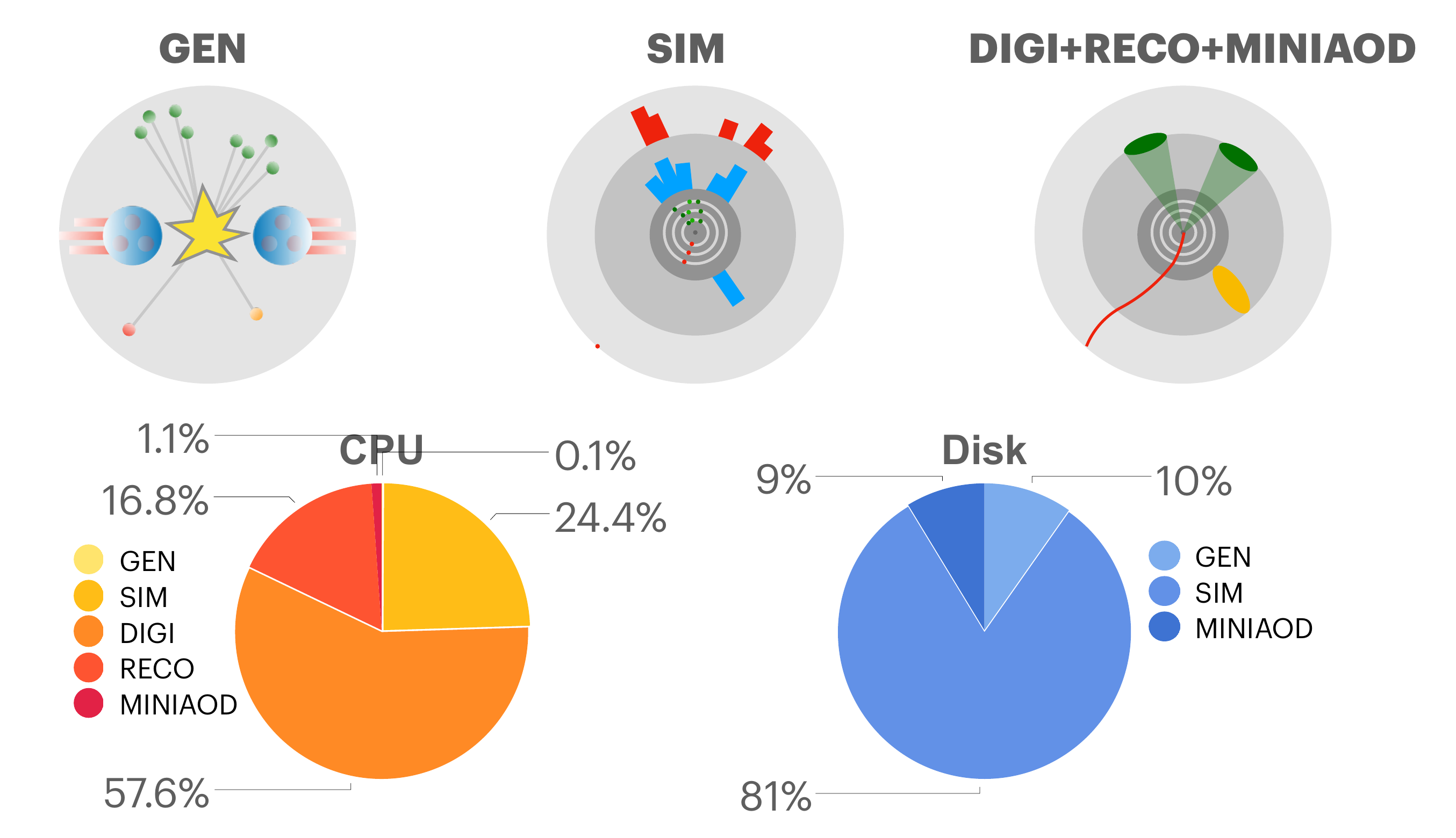}
    \caption{TOP: The event generation workflow of the CMS experiment. The proton-proton collision process is simulated up to the production of stable (hence observable) particles (GEN). The simulation of the detector response is modelled by the {\tt GEANT4} library (SIM). The resulting energy deposits are turned into digital signals (DIGI) that are then reconstructed by the same software used to process real collision events (RECO). At this stage, high-level objects such as jets are reconstructed. Starting from the RECO data format, a reduced analysis data format (MINIAOD) is derived. BOTTOM: computing resource breakdown for the generation workflow of the CMS experiment, in terms of CPU (left) and storage disk (right). See Appendix~\ref{app:CMSSW_workflow} for details.}
    \label{fig:cms_gen}
\end{figure}

To give a concrete example, we consider the event simulation workflow of the CMS experiment, schematically represented in Fig.~\ref{fig:cms_gen}. The first step (GEN) consists in running an event generator library, simulating a proton-proton collision, the production of high-mass particles from it, and the decay of these particles to those stable particles which are then seen by the detector. This step creates the so-called generator-level view of a collision event, corresponding to what a perfect detector would see. The simulation of the detector response (SIM) translates this flow of particles into a set of {\it detector hits}, taking into account detector imperfections and the limited experimental resolution. These hits are converted to the same digital format (DIGI) produced by the detector electronics and then reconstructed by the same software used to process real collision events (RECO). At this stage, high-level objects such as jets are created. Starting from the RECO data format, a reduced analysis data format (MINIAOD) is derived~\cite{Petrucciani:2015gjw}. Figure~\ref{fig:cms_gen} also provides a breakdown of CPU and disk resources for each of these steps. Details on the procedure followed to measure these values are given in Appendix~\ref{app:CMSSW_workflow}.

Recently, generative algorithms based on Deep Learning (DL) techniques have been proposed as a possible solution to speed up  {\tt GEANT4}. When following this approach, one typically focuses on an image representation
of LHC collisions (e.g., energy deposits in a calorimeter) and develops some kind of generative model~\cite{Goodfellow:2014upx,arjovsky2017wasserstein,1704.00028,rezende14,kingma2014auto} to by-pass {\tt GEANT4} when simulating the detector response to individual particles~\cite{Paganini:2017dwg,Erdmann:2018jxd,Salamani:2645142,Belayneh:2019vyx,Buhmann:2020pmy} or to groups of particles, such as jets~\cite{deOliveira:2017pjk,Musella:2018rdi,Carrazza:2019cnt} or cosmic rays~\cite{Erdmann:2018kuh}. Generative models were considered also for similar applications in HEP, such as amplitude~\cite{matrix_ml} and full event topology~\cite{DiSipio:2019imz,Butter:2019cae,Martinez:2019jlu} generation. While these studies demonstrate the potential of generative models for HEP, more work is needed to fully integrate this new methodology in the centralized computing system of a typical LHC experiment. In particular, one needs to work beyond the collision-as-image paradigm so that the DL-based simulation accounts for the irregular geometry of a typical detector while delivering a  dataset in a format compatible with downstream reconstruction software. 

Other studies~\cite{Otten:2019hhl,Hashemi:2019fkn} investigated a more extreme approach: rather than training models to perform generic generation tasks in a broader software framework (e.g., a DL-based shower generator in GEANT), one could design analysis-specific generators, with the limited scope of delivering arrays of values for physics quantities which are relevant to a specific analysis. Reducing the event representation to a vector of meaningful quantities, one could obtain a large amount of events in short time  and with small storage requirements by skipping all the intermediate steps of the data processing. The considered features could be the fundamental quantities used by a given analysis (e.g., the four-momenta of the final-state reconstructed objects in a search for new particles). In this context, both generative adversarial networks (GANs)~\cite{Otten:2019hhl,Hashemi:2019fkn} and variational autoencoders (VAEs)~\cite{Otten:2019hhl} were considered. In this case, one learns the N-dimensional probability density function (N-dim pdf) of the event, in a space defined by the quantities of interest for a given analysis. Sampling from this function, one can then generate new data. The open question with this approach stands with the trade-off between statistical precision (which decreases with the increase amount of generated events) and the systematic uncertainty that could be induced by a non accurate description of the N-dim pdf. When training both VAEs and GANs, one learns how to interpolate between the samples provided in the training dataset. The limited amount of data in the training dataset is the ultimate precision-limiting factor, as discussed in Ref.~\cite{matchev}, but generative models retain amplification capability similarly to what a fitting function does, as shown in Ref.~\cite{butter2020ganplifying} for GANs. Ultimately, one needs to balance the statistical uncertainty (i.e., the amplification factor when augmenting the dataset) and systematic uncertainties associated to the accuracy with which the generative model interpolates between the training data points. The balance will be reached tuning, among other things, the training dataset size. The optimal configuration, intrinsically application specific, determines whether a generative model is computationally convenient.\footnote{Here, we are assuming that {\tt GEANT4} will be used to generate the training dataset and the generative model will then be used to scale up the simulated dataset size.  If the desired accuracy can be reached only at the price of more training data to be generated, the net gain of this approach would be reduced.}

In this paper, we propose to rephrase the problem of analysis-specific dataset generation. Rather than morphing a distribution in a latent space into a target distribution, we want to start from the ideal-detector distribution and morph it into the actual-detector distribution, learning a fast-and-accurate detector response model. We do so combining the strength of multi-dimensional deep neural regressors to the adaptive power of kernel density estimation, which has a long and successful tradition in particle physics~\cite{Cranmer:2000du}. For a given physics study, we assume that the interesting features can be represented by a limited set of high-level quantities (the feature vector $\vec{x}$). We assume that a training dataset is provided. For each collision event in the dataset, the feature vector is computed at three stages: (i) {\it at generator level} $\vec{x}_G$, i.e., before applying any detector simulation. This view of the collision event corresponds to the perfect-resolution ideal detector case; (ii) {\it at reconstruction level} $\vec{x}_R$, i.e. after the simulation of the detector response, modelled with {\tt GEANT4}; (iii) at the output of the DL model $\vec{x}_{DL}$. We model the detector response as a function of the generator-level feature vector:
\begin{equation}
x^i_{DL} =  {\cal N}(\mu^i_R(\vec{x}_G), \sigma^i_R(\vec{x}_G))~, 
\label{eq:genModel}
\end{equation}
where ${\cal N}(\mu, \sigma)$ is a one-dimension Normal function centered at $\mu$ with variance $\sigma^2$ and the index $i$ runs over the components of the feature vector $\vec{x}$. We train a DL model to simultaneously learn the functions $\vec{\mu}_R(\vec{x}_G)$ and $\vec{\sigma}_R(\vec{x}_G)$ and then use the Normal model of Eq.~(\ref{eq:genModel}) to generate $\vec{x}_{DL}$ from $\vec{x}_G$. Under the assumption that large sets of $\vec{x}_G$ values can be obtained in relatively short time (which is typically the case for High Energy Physics applications), this strategy would result in a sizable save of computing resources. On one hand, one would reduce computing time bypassing the more intense steps of the generation workflow. In addition, one would reduce the need for large storage elements: rather than storing individual collision data, which demands an event storage allocation between ${\cal O}(1 \mathrm{MB})$ (for raw data) and ${\cal O}(10 \mathrm{kB})$ (for analysis-ready object collections), one would directly handle a few relevant quantities for a given analysis. 
One could save resources by utilizing analysis-specific fast simulation models for data augmentation, e.g., generating 10\% of the required data with the traditional {\tt GEANT4} workflow and the remaining 90\% only up to the GEN step. These data, shared among the ${\cal O}(100)$ analyses, would be used to create analysis-specific training and inference datasets. Even considering that ${\cal O}(100)$ analysis teams would have to train ${\cal O}(100)$ specific generative models, the strategy we propose would result in an important resource gain, provided a large enough training facility.\footnote{The model presented in this work was trained on a RTX2080 GPU by NVIDIA in 30 minutes. Even a small-size dedicated GPU cluster with O(10) GPUs dedicated to this use case could then serve the needs of a large collaboration. Its cost is negligible on the scale of the large computing infrastructures built for the LHC experiments.}

We demonstrate this strategy at work on a concrete example, namely the generation of $W$+1~jet events produced in $\sqrt{s}=$~13~TeV proton-proton collisions, similar to those recorded at the LHC. We discuss the model design and training, its performance and its accuracy for factor-ten data augmentation. 

This paper is structured as follows: Section~\ref{sec:data} provides a full description of the input dataset and its feature-vector representation. Section~\ref{sec:architecture} describes the model architecture and the training setup. Sections~\ref{sec:results}~and~\ref{sec:resources} discuss the model performance in terms of accuracy and resource utilization, respectively. Conclusions and outlook are given in section~\ref{sec:conclusions}.

\section{Benchmark dataset}
\label{sec:data}

As a benchmark problem, we consider the generation of $W$+1~jet events produced in $\sqrt{s}=$~13~TeV proton-proton collisions. The starting point is the inclusive production of $W\to \mu \nu$ events using {\tt PYTHIA8}~\cite{pythia}. At this stage, we require each event to have at least one muon with a transverse momentum $p_T>22$~GeV.\footnote{We use a Cartesian coordinate system with the $z$ axis oriented along the beam axis, the $x$ axis on the horizontal plane, and the $y$ axis oriented upward. The $x$ and $y$ axes define the transverse plane, while the $z$ axis identifies the longitudinal direction. The azimuth angle $\phi$ is computed with respect to the $x$ axis. The polar angle $\theta$ is used to compute the pseudorapidity $\eta = -\log(\tan(\theta/2))$. The transverse momentum ($p_T$) is the projection of the particle momentum on the ($x$, $y$) plane. We fix units such that $c=\hbar=1$.} Detector effects are modelled using {\tt DELPHES} v3.4.2. We consider the CMS detector model for the HL-LHC  upgrade, distributed with {\tt DELPHES}. At this stage, the event is overlaid to 
minimum-bias events to model the effect of pileup, i.e., those parasitic proton-proton collisions happening at the same beam crossing as the interesting event. For each collision, the number of pileup collisions is sampled from a Poisson distribution with expectation value set at 200, in order to match the expected conditions for HL-LHC. 

At generator level (GEN), jets are clustered using the {\tt anti-kt} algorithm~\cite{antiKT} with jet-size parameter $R=0.5$, taking the four-momenta of all the stable particles in the event as input. We consider events with one clustered jet, with $p_T>30$~GeV and $|\eta|<2.4$. In order to avoid the double counting of muons as jets, we require
$\Delta R = \sqrt{\Delta \eta^2 + \Delta \phi^2}>0.5$ between the muon and the jet in each event. 

At reconstruction level, jets are clustered from the list of particles returned by the {\tt DELPHES} particle-flow algorithm. As for the GEN jets, we consider {\tt anti-kt} jets with $R=0.5$. Both the muon and jet are matched to the corresponding generator-level object. In addition, we discard mismatched muons by requiring that the relative residual of the muon $p_T$ to be $|p_T^{G}-p_T^R|/p_T^G<10\%$.

The feature vector $\vec{x}$ is built considering the following nine quantities:
\begin{itemize}
    \item The muon momentum in Cartesian coordinates: $p_x^\mu$, $p_y^\mu$, and $p_z^\mu$.
    \item The jet momentum in Cartesian coordinates: $p_x^j$, $p_y^j$, and $p_z^j$.
    \item The logarithm of the jet mass $\log(M_j)$.
    \item The missing transverse energy in Cartesian coordinates: $E^{\mathrm{miss}}_x$ and $E^{\mathrm{miss}}_y$.
\end{itemize}

In addition, we consider a set of 12 auxiliary features, computed from
the input feature vector $\vec{x}$:
\begin{itemize}
\item The muon momentum in longitudinal-boost-invariant coordinates: $p_T^\mu$, $\eta^\mu$, and $\phi^\mu$.
\item The jet momentum in longitudinal-boost-invariant coordinates: $p_T^j$, $\eta^j$, and $\phi^j$.
\item The missing transverse energy in polar coordinates: $E^{\mathrm{miss}}_T$ and $\phi_{\mathrm{miss}}$.
\item The transverse mass $M_T$, i.e., the mass of the four momentum obtained summing the the muon transverse momentum $(E_T^\mu, p_x^\mu, p_y^\mu, 0)$ to the missing transverse energy $(E^\mathrm{miss}_T, E^\mathrm{miss}_x, E^\mathrm{miss}_y, 0)$.
\item $S_T$, i.e., the scalar sum of $E^\mathrm{miss}_T$, $p_T^\mu$, and $p_T^j$.
\item The jet mass: $M_j$.
\end{itemize}
These quantities are computed at generator and reconstruction level and are used to assess how well the correlation between the generated quantities is modeled. Unlike the feature-vector quantities, they do not enter the definition of the loss function.

The model training and performance assessment is done on a dataset of 2M events, which we separate in a test and a learning datasets, containing 20\% and 80\% of the events, respectively. The learning dataset is further split into a training (70\%) and a validation (30\%) dataset. In order to test the data augmentation properties of the proposed strategy, we also consider a larger test dataset, containing 10M events. 

\section{Model description and training}
\label{sec:architecture}

\begin{figure}[t!]
    \centering
    \includegraphics[width=0.95\textwidth]{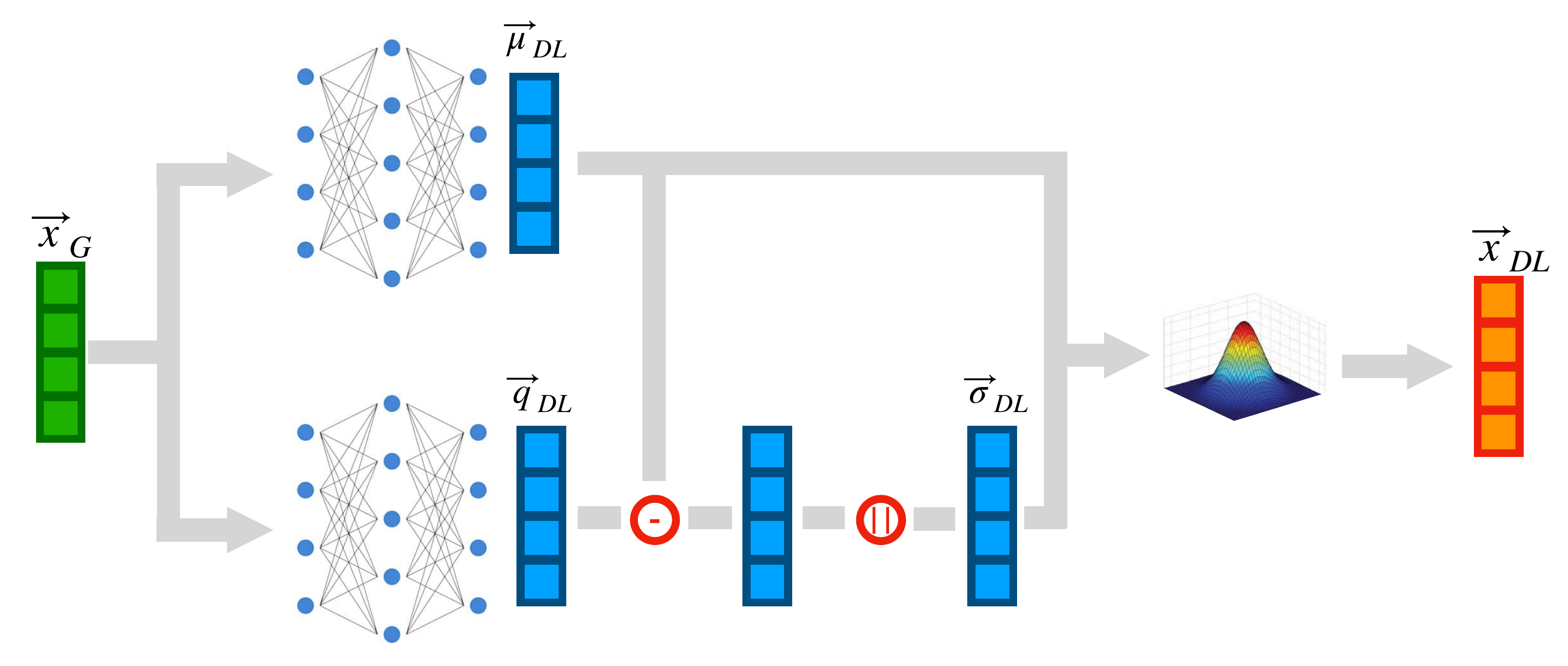}
    \caption{Model architecture: a feature vector at generator level $\vec{x}_G$ is given as input to two regression models, returning vectors of central values ($\vec{\mu}_R$) and RMS ($\vec{\sigma}_R$), from which the reconstructed feature vector predicted by the DL model $\vec{x}_{DL}$ is generated.\label{fig:model_arch}}
\end{figure}

Our model architecture is represented in Fig.~\ref{fig:model_arch}. The input vector $\vec{x}_G$ of generator-level features is passed to two regressive models, each returning a vector with the same dimensionality of $\vec{x}_G$. One is interpreted as a vector of mean values $\vec{\mu}_{DL}$.
The other one is interpreted as the "$\pm 1\sigma$" quantile $\vec{q}_{DL}$. By taking the absolute difference between each mean value and its corresponding quantile, we compute the RMS values $\vec{\sigma}_{DL}$.

Each regressive model consists of a six-layer dense neural network. The first and last layers have nine nodes each, while the intermediate layers have 100 nodes. All layers except the last one are activated by LeakyReLU~\cite{Maas13rectifiernonlinearities} functions, with $\alpha=0.05$. Linear activation functions are used for the last layer. 
The model output is then computed as $\vec{x}_{DL} = {\vec \mu}_{DL} + \vec{\sigma}_{DL} \cdot \vec{\epsilon}$, where the vector $\vec{\epsilon}$ contains random numbers sampled from a Normal function centered at 0 with unit variance. In addition to the main features $\vec{x}$, we compute a set of auxiliary features (see section~\ref{sec:data}) used for a further post-training validation.

We define the output of the DL model $\vec{x}_{DL}$ to the $\vec{x}_{R}$ returned by the simulation. The loss function is written as the sum of a mean estimate and a quantile regression:
\begin{equation}
    {\cal L}_{RECO} = \frac{1}{N} \sum_{i=1}^N \left [ \sum_{j=1}^k | \mu_{DL}^{i,j} - x_{R}^{i,j} | + \Theta(q_{DL}^{i,j}, x_{R}^{i,j}) | q_{DL}^{i,j} - x_{R}^{i,j} | \right ]~,
    \label{eq:Lreco}
\end{equation}

where the sum on $i$ runs over the training dataset with cardinality $N$ and  $j$ over the $k$ dimensions of $\vec{x}$. The function $\Theta (x, y)$ is defined as:
\begin{equation}
    \Theta (x, y) = (1-\gamma) \theta(x-y) + \gamma \theta(y-x)~,
\end{equation}
where the step function $\theta(t)$ is set to one (zero) for positive (negative) values of $t$ and $\gamma = 0.841$. This choice of $\gamma$ guarantees that the loss is minimized to learn the quantile corresponding to one standard deviation. 

  

We implement the model in {\tt KERAS}~\cite{keras} and train it with the Adam~\cite{adam} optimizer, with batches of 128  and an epoch-dependent learning rate $lr = 0.001/(1+n_\mathrm{epoch})$. The model is trained for 100 epochs, but convergence is typically reached between 30 epochs.
The network parameter values corresponding to the smallest validation loss are taken as the optimal configuration.

\section{Results}
\label{sec:results}

The trained model is used to generate samples of reconstructed events from generator-level events. We evaluate the training performance by comparing the output distributions with those obtained by {\tt DELPHES} for the same generator-level events. 

\begin{figure}[t!]
    \centering
    \includegraphics[width=0.8\textwidth]{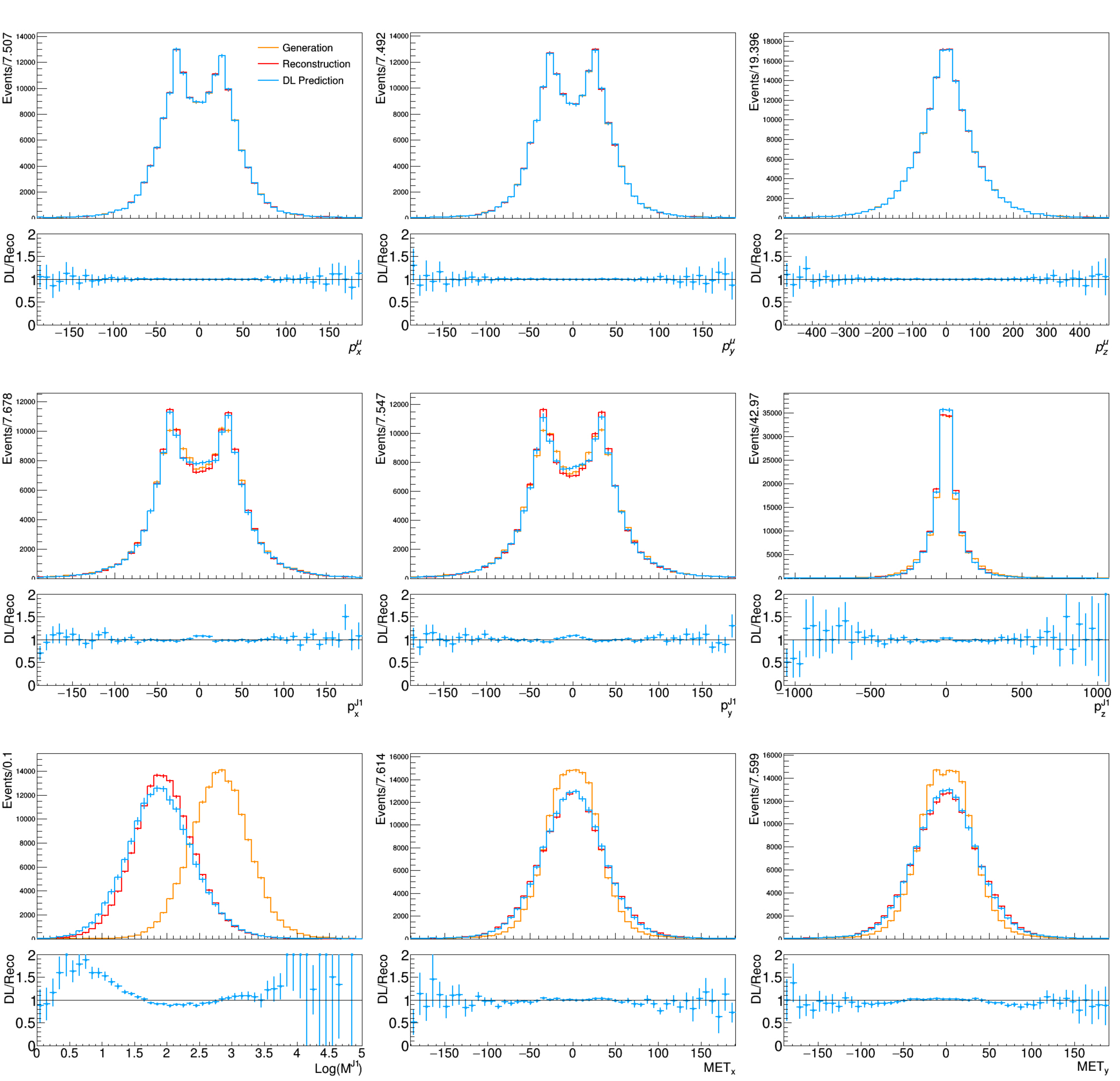}
    \caption{Distribution of reconstructed and model-predicted quantities for the feature-vector quantities, 
    compared to the corresponding quantities from generator-level quantities provided as input to the model. \label{fig:dist_input}}
\end{figure}

\begin{figure}[t!]
    \centering
    \includegraphics[width=\textwidth]{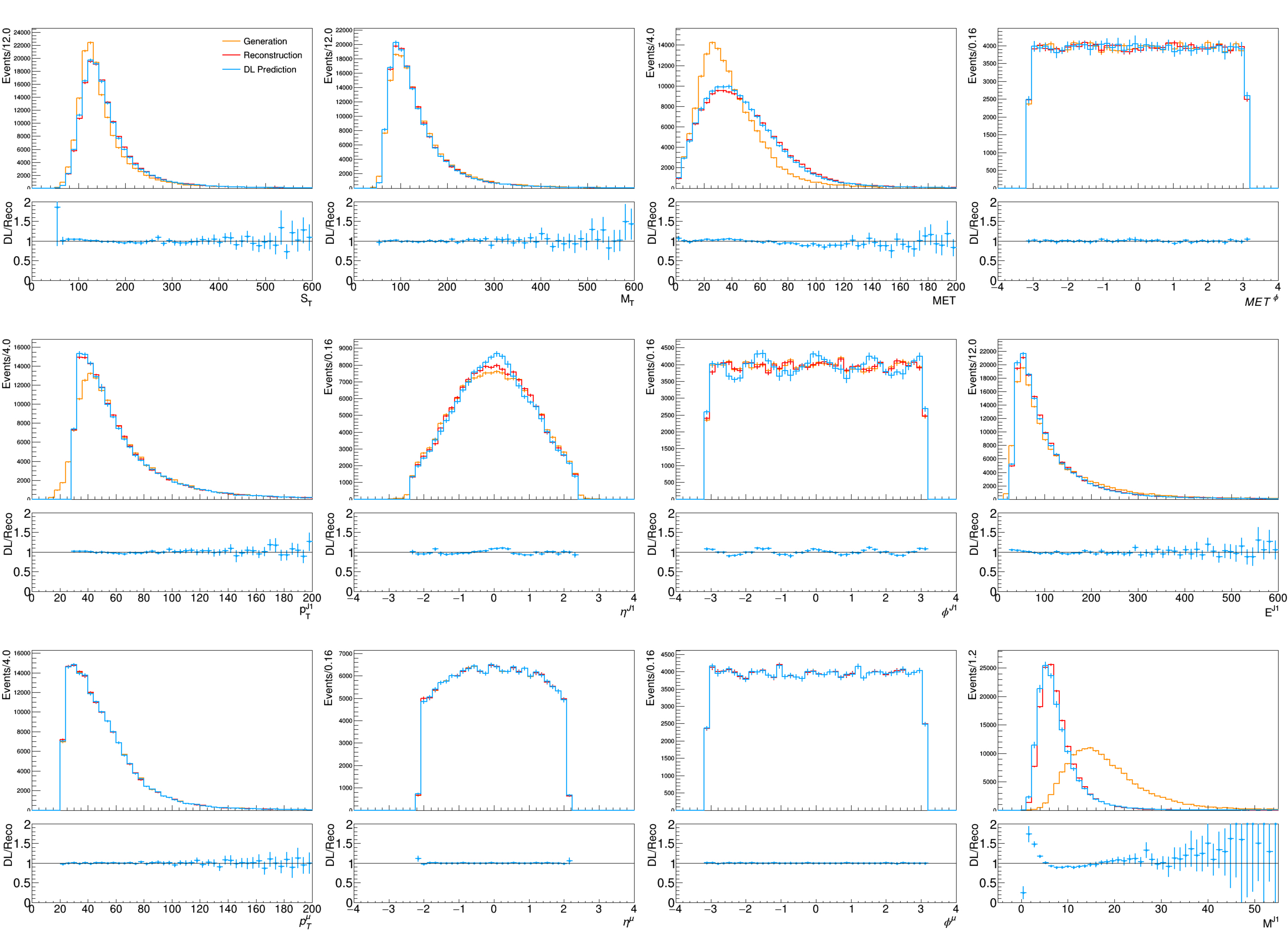}
    \caption{Distribution of reconstructed and model-predicted auxiliary quantities, 
    compared to the corresponding generator-level quantities. \label{fig:dist_aux}}
\end{figure}

A comparison is shown in Fig.~\ref{fig:dist_input} for the feature-vector quantities. The sample derived from the DL model is  similar to the the one obtained running a classic generation workflow. The model can account for small perturbations and major distortions of the GEN distribution, as well as the default detector simulation workflow. The agreement is not perfect, and certainly the model can be improved. Nevertheless, the reached accuracy is comparable to that of a typical data-to-simulation comparison and certainly sufficient to support the novel procedure that we want to put forward in this study. The observed agreement goes beyond one-dimensional projections of the input features. The distributions of auxiliary quantities, computed as a function of the feature-vector quantities, are also modelled to a good precision (see Fig.~\ref{fig:dist_aux}). This demonstrates that the DL-based generator accounts for correlations between quantities, as much as the traditional {\tt DELPHES}-based workflow does.

\begin{figure}[t!]
    \centering
    \includegraphics[width=0.8\textwidth]{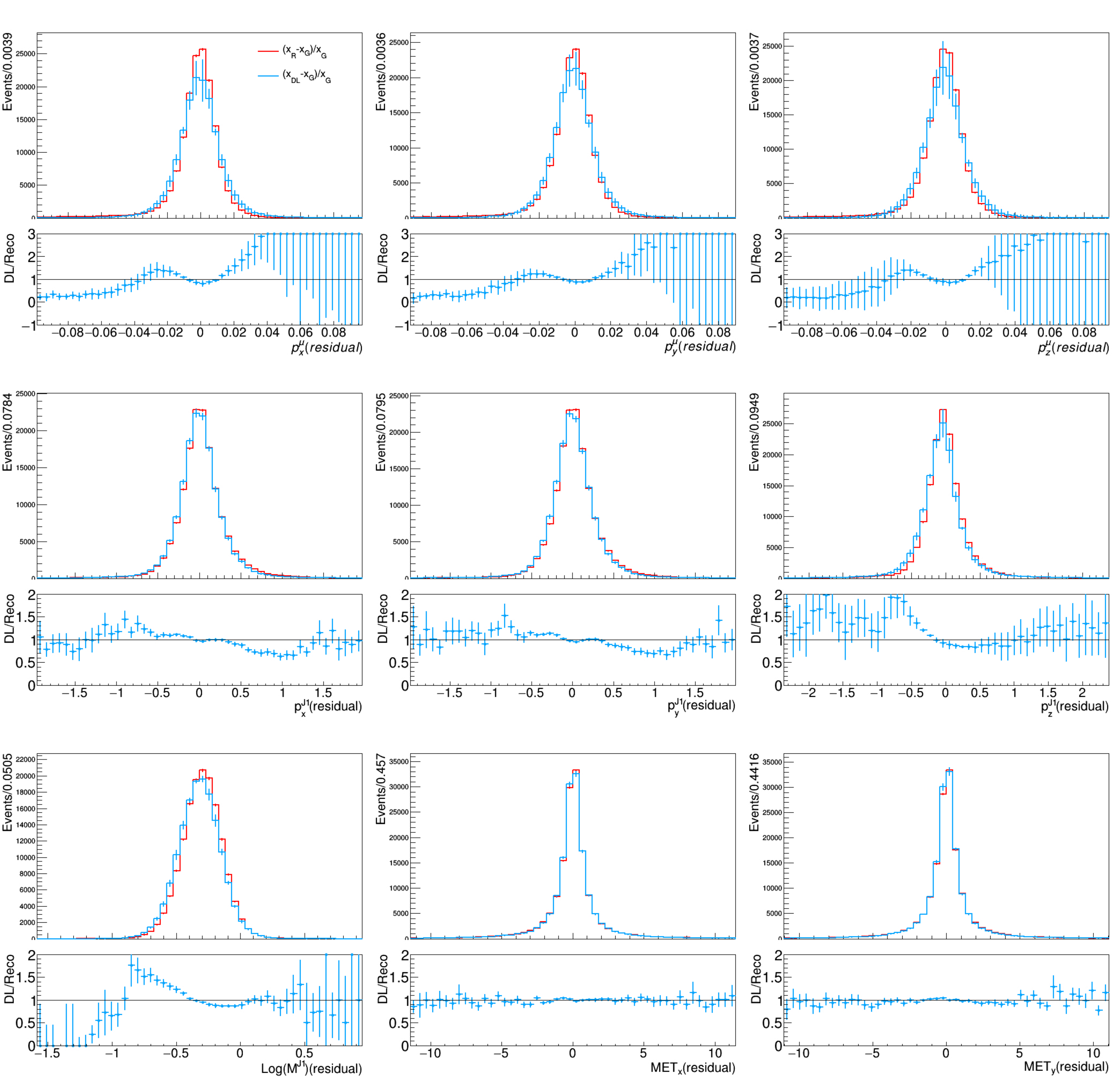}
    \caption{Relative residual distribution for reconstructed and model-predicted quantities in the feature vector, computing with respect to the reference input. The bottom panel of each plot shows the ratio between the two relative residuals, expected to be consistent with $1$ for a DL model which correctly models the detector response of the traditional workflow. \label{fig:residual_input}}
\end{figure}

\begin{figure}[t!]
    \centering
    \includegraphics[width=\textwidth]{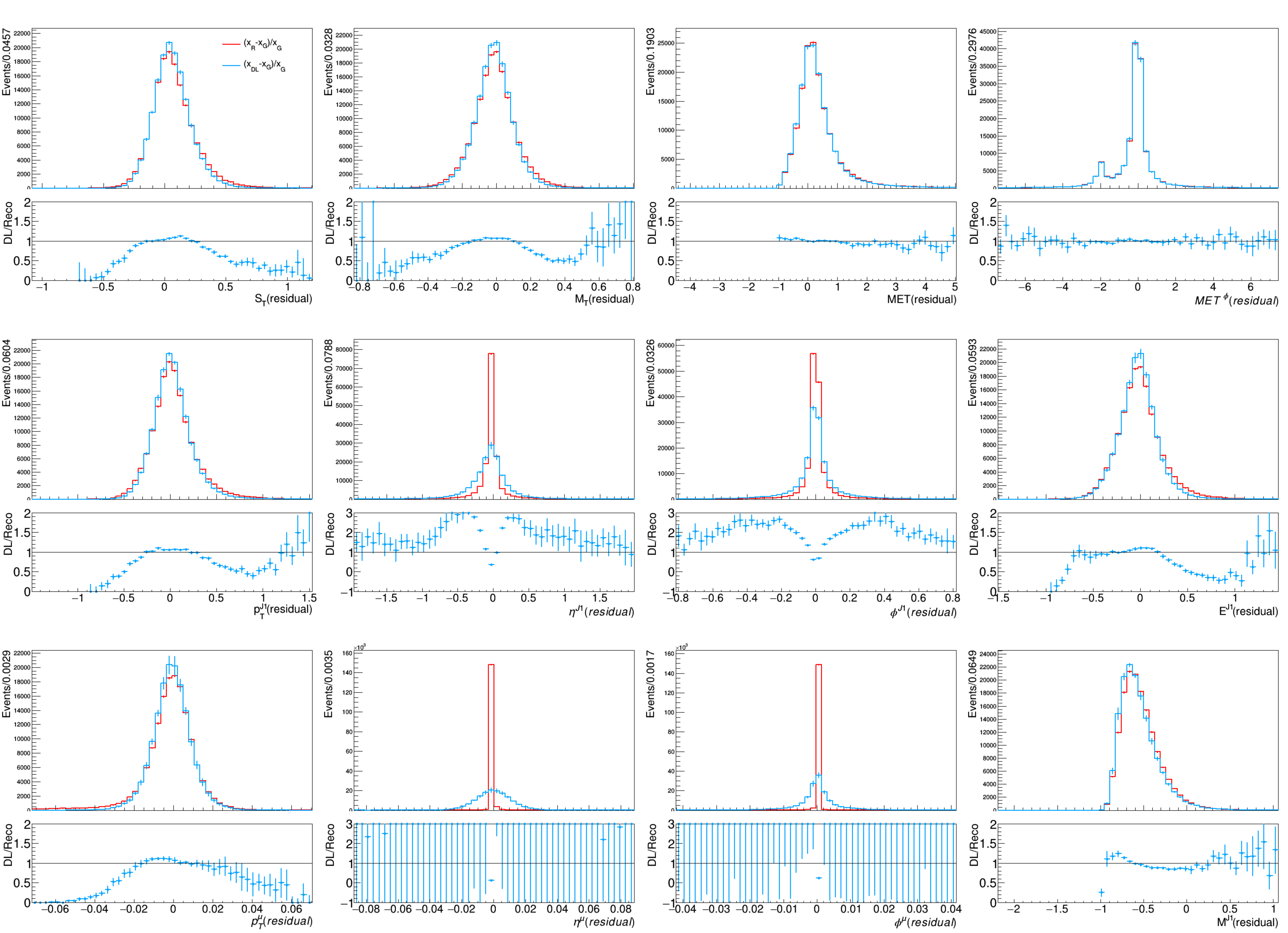}
    \caption{Relative residual distribution for reconstructed and model-predicted auxiliary quantities, computing with respect to the reference input. The bottom panel of each plot shows the ratio between the two relative residuals, expected to be consistent with $1$ for a DL model which correctly models the detector response of the traditional workflow.\label{fig:residual_derived}}
\end{figure}

While a comparison of dataset distribution gives a confidence of the quality achieved by the DL model, one can further test the achieved precision by looking at relative residual distributions. Our DL model does not sample events from a latent space (like a GAN or a plain VAE). Instead, it  works as a fast simulation of a given generator-level event, preserving the correspondence between the reconstructed and the generated event, which allows us to compare event-by-event relative residual distributions. These distributions, which quantify the detector effects on the analysis-specific interesting quantities, are shown in Fig.~\ref{fig:residual_input}. There, we compare the relative residuals between reconstructed and generated quantities, for the DL-based and the traditional simulation workflow. In this case, we train the model ten times and produce ten relative residual distributions. The bin-by-bin spread of these distributions is considered as a systematic uncertainty associated to the DL model, which is summed in quadrature to the statistical uncertainty in the same bin to compute the total uncertainty, shown by the error bars of the DL model in the figure. Only the statistical uncertainty is shown for the corresponding distributions of reconstructed quantities. An overall agreement is observed, despite a small bias on the muon momentum coordinates. Figure~\ref{fig:residual_derived} shows the same comparison for the auxiliary quantities. As the plot shows, a correct modeling of the residuals is obtained for energies, masses, and momenta. On the other hand, the model struggles to model the high-resolution detector response on the $\eta$ and $\phi$ coordinates. While this has little impact on the modeling of the $\phi$ and $\eta$ distributions (see Fig.~\ref{fig:dist_input}), this is certainly an aspect to improve in real-life applications.

Appendix~\ref{app:2Ddist} provides further assessments of the generation quality, showing 2D distributions of quantities derived from the DL-based generator vs the traditional one.

\begin{figure}
    \centering
    \includegraphics[width=0.8\textwidth]{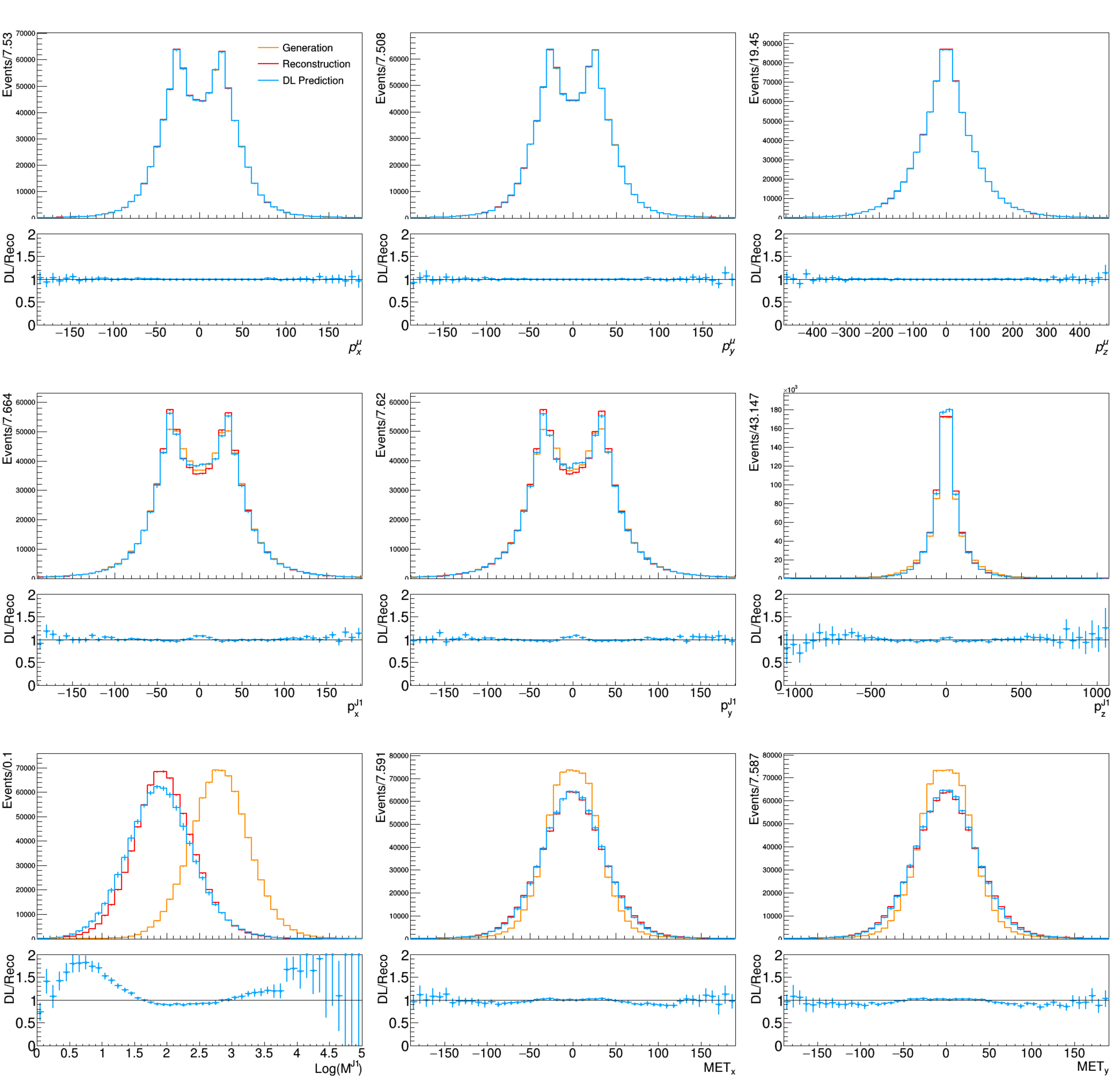}
    \caption{Distribution of reconstructed and model-predicted quantities for the feature-vector quantities, 
compared to the corresponding quantities from generator-level input. In this case, the model is applied to a dataset five times larger than the training dataset.\label{fig:dist_input_more}}
\end{figure}

\begin{figure}
    \centering
    \includegraphics[width=\textwidth]{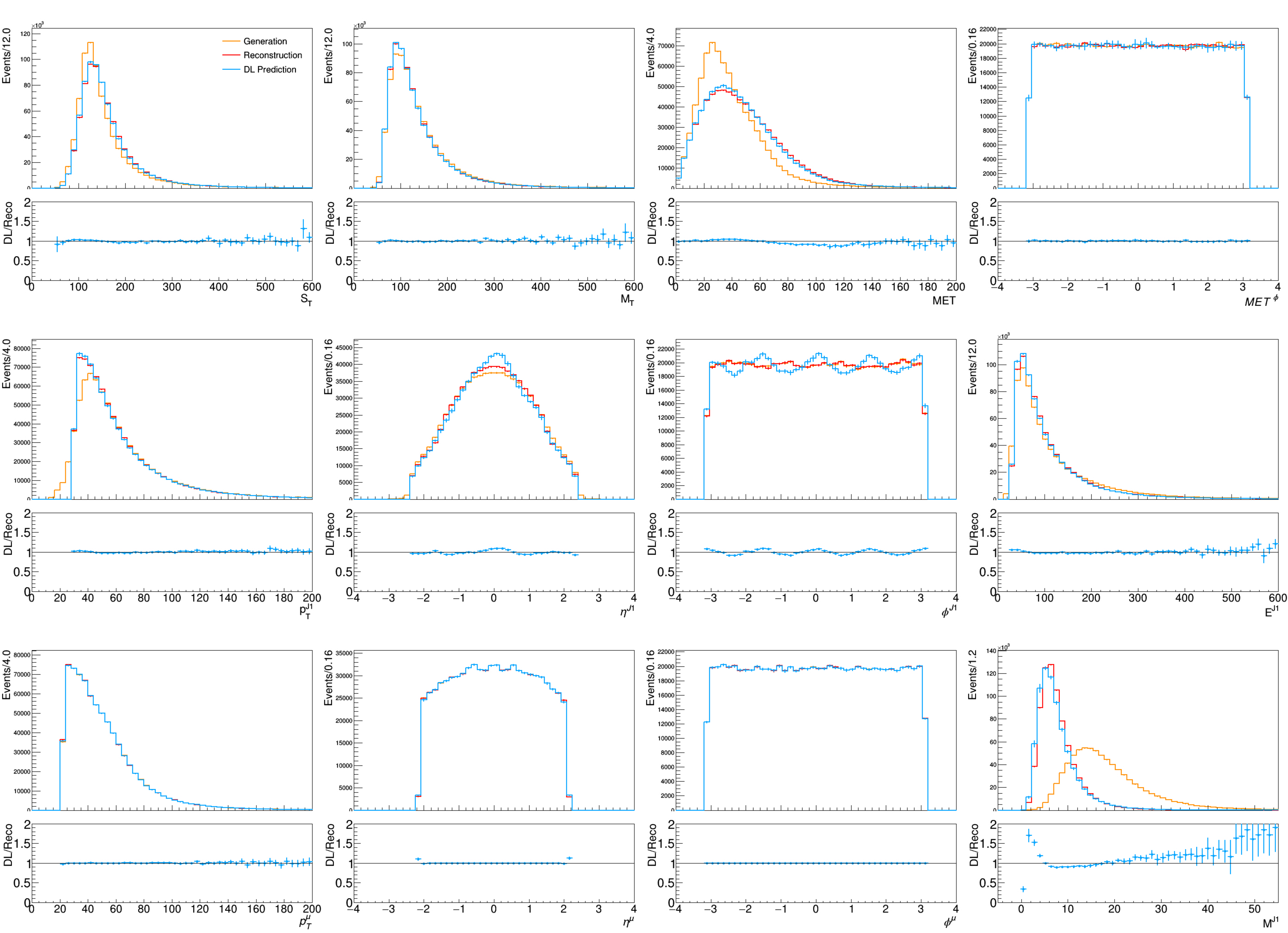}
    \caption{Distribution of reconstructed and model-predicted quantities in the auxiliary quantities, 
    compared to the corresponding quantities from generator-level input. In this case, the model is applied to a dataset five times larger than the training dataset.\label{fig:dist_derived_more}}
\end{figure}

In order to test the scaling of model accuracy with the inference dataset size, we apply our DL-based fast simulation strategy to a dataset five times bigger than what used for training. 
Figures~\ref{fig:dist_input_more}~and~\ref{fig:dist_derived_more} show the comparison of the distributions obtained in this case, compared to what is obtained with {\tt DELPHES}, respectively for the input vector and the auxiliary features. The corresponding relative residual distributions are shown in Appendix~\ref{app:five_time_stat}. These distributions agree with those obtained when the training and inference dataset size agree, i.e., no accuracy deterioration is observed due to the scaling of the dataset size. This fact proves the robustness of the proposed methodology and its effectiveness as a data augmentation tool.

\section{Computing resources}
\label{sec:resources}

In order to fully assess the advantage of the proposed generation workflow, we consider the following use case: an analysis team requests $N$ events to be centrally produced by the central computing infrastructure of their experimental collaboration. Instead, the central system would deliver $N$ events at generator level (GEN step of Fig.~\ref{app:CMSSW_workflow}), while processing only $n<N$ of them through the full chain. The analysis team would then (i) run their data analysis software on the $n$ events, and (ii) train on these data a DL-based fast-simulation like the one presented in Section~\ref{sec:architecture}. With this model, they would then (iii) process the other $(N-n)$ generator-level events and produce the dataset required for their analysis. 

In order to assess the resource savings, we point out that step (iii) comes with negligible computational costs. Model inference on a CPU requires 100 sec to run on 100000 events (i.e., ${\cal O}(1)$~msec/event), which results in a 8 MB file (saved as a compressed HDF5 file) for the example use case we discussed. While these details would change depending on the analysis-specific event representation, the quoted values give a reasonable order-of-magnitude estimate of the expected resource needs.
Step (ii) can run at a minimal cost: our model could train within 30 minutes when running on a commercial GPU. The residual cost is then entirely driven by step (i). While a traditional workflow requires O(100) sec/event of CPU time and occupies O(1) MB/event of storage, producing the same statistics ($N$ events) of GEN-only events would require 10\% disk allocation with a negligible CPU cost, as shown in Fig.~\ref{app:CMSSW_workflow}. 

As a consequence, by adopting the strategy outlined above, one would save a factor $N/n$ in CPU (i.e., only spend sizable CPU resources to produce the training dataset, which would remain generic and could serve more than one analysis). The storage allocation would result from the sum of $n$ events in full format and $(N-n)$ GEN-only events, for a total saving of $N/(n + 10\% (N-n))$. For instance, considering $N=1M$ events and $n=10\% N$, one would save $90\%$ of the CPU resources and $79\%$ of the disk storage, almost equally shared among the full-format training data and the $(N-n)$ GEN-only data. 

\section{Conclusions}
\label{sec:conclusions}

We presented a proposal for a new data augmentation strategy for fast simulation workflows at LHC experiments, which exploits a generative Deep Learning model to convert an analysis-specific representation of collision events at generator level to the corresponding representation at reconstruction level. Following this procedure, one could replace any request of $N$ simulated events with an $n<N$ request, providing the residual $(N-n)$ events at generator level. Bypassing the detector simulation and reconstruction process for the  $(N-n)$ events, one would benefit of a substantial reduction in terms of required resources. 

We demonstrated that a simple mean-and-variance regression
model with a Gaussian sampling function allows to reach a good performance, producing a dataset which resembles that from a traditional workflow. We showed that the accuracy is preserved when applying our strategy to a test dataset much larger than the training dataset.

The proposed model is much simpler than a generative model, e.g., a GAN. The architecture is easier to train and the task it learns to solve is simpler than generative realistic events from random points in a latent space. The generator-level input carries much of the domain knowledge and the statistical fluctuations of the target dataset size. In addition, thanks to the light computational weight of the training and inference steps, one could consider to train several models and apply them to the same test dataset, using the spread of predictions to evaluate a simulation systematic uncertainty. 

We believe that the LHC experiments could benefit from adopting the proposed procedure, particularly for the high-precision measurement era during the High-Luminosity LHC phase.

\section*{Acknowledgments}


This project is partially supported by the European Research Council
(ERC) under the European Union's Horizon 2020 research and innovation
program (grant agreement n$^o$ 772369) and by the United States
Department of Energy, Office of High Energy Physics Research under
Caltech Contract No. DE-SC0011925.

This work was conducted at ``\textit{iBanks},'' the AI GPU cluster at
Caltech.  We acknowledge NVIDIA, SuperMicro and the Kavli Foundation
for their support of ``\textit{iBanks}.''

\section*{Appendix}

\appendix

\section{Resource utilization for a standard GEANT4-based generation workflow}
\label{app:CMSSW_workflow}

In this appendix, we describe how we derived the values quoted in Fig.~\ref{fig:cms_gen}. We take as a reference the CMS experiment. In absence of a published reference with a breakdown of CPU and disk resources for GEN, SIM, and DIGI+RECO steps, we derived the quantities quoted in Fig.~\ref{fig:cms_gen} by generating QCD events on CPU, through the CERN batch system.
To do so, we relied on the open-source CMSSW software~\cite{cmssw_github} and followed the instructions provided by the CMS collaboration on the CERN Open Data portal~\cite{cms_opendata}. 

We consider the same setup used to generate one of the QCD Run II samples published on the CERN Open Data portal~\cite{QCD_Pt_470to600} and the software installation available on CERN {\tt cvmfs} distributed file system. 

For each step, we ran jobs with 100 and 10 events. For each job, we recorded CPU time and output file size. Each step is repeated 10 times and the average of each quantity is considered. The typical uncertainty on these mean values, measured by the standard deviation of the 10 values, is found to be at most of a few percent and hence considered negligible. 
After computing the average for each set of jobs, we take the difference between the 100-event and 10-event job of each kind, in order to remove the overhead CPU time and file size that doesn't originate from per-event tasks. 
By dividing these differences by 90, we derive the per-event quantities quoted in Fig.~\ref{fig:cms_gen}.

\section{Further validation of the DL generation workflow}
\label{app:2Ddist}

\begin{figure}[t!]
    \centering
    \includegraphics[width=0.8\textwidth]{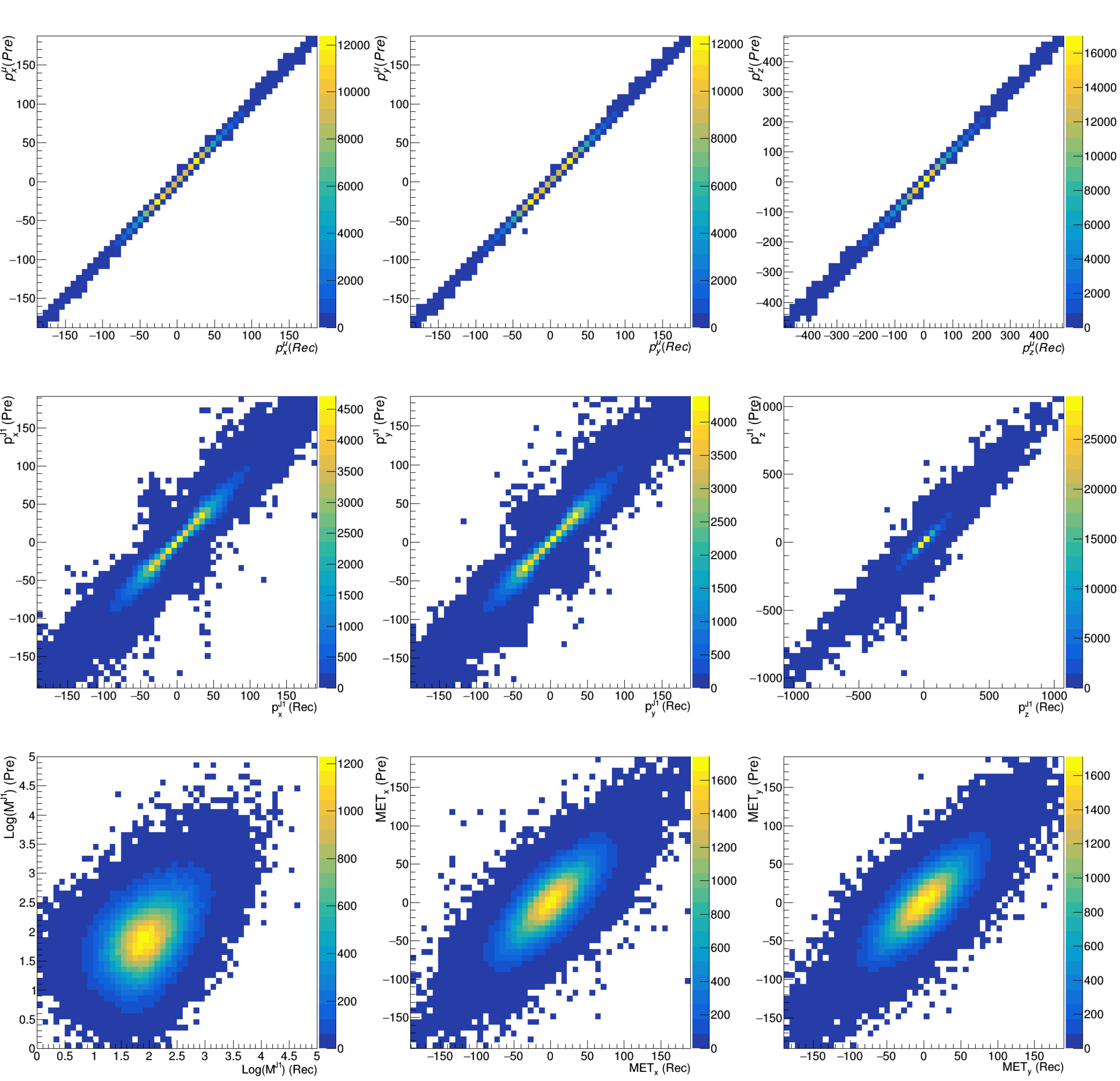}
    \caption{Distribution of input features predicted by the model as a function of the corresponding quantities from detector simulation.\label{fig:2D_input}}
\end{figure}

\begin{figure}[t!]
    \centering
    \includegraphics[width=\textwidth]{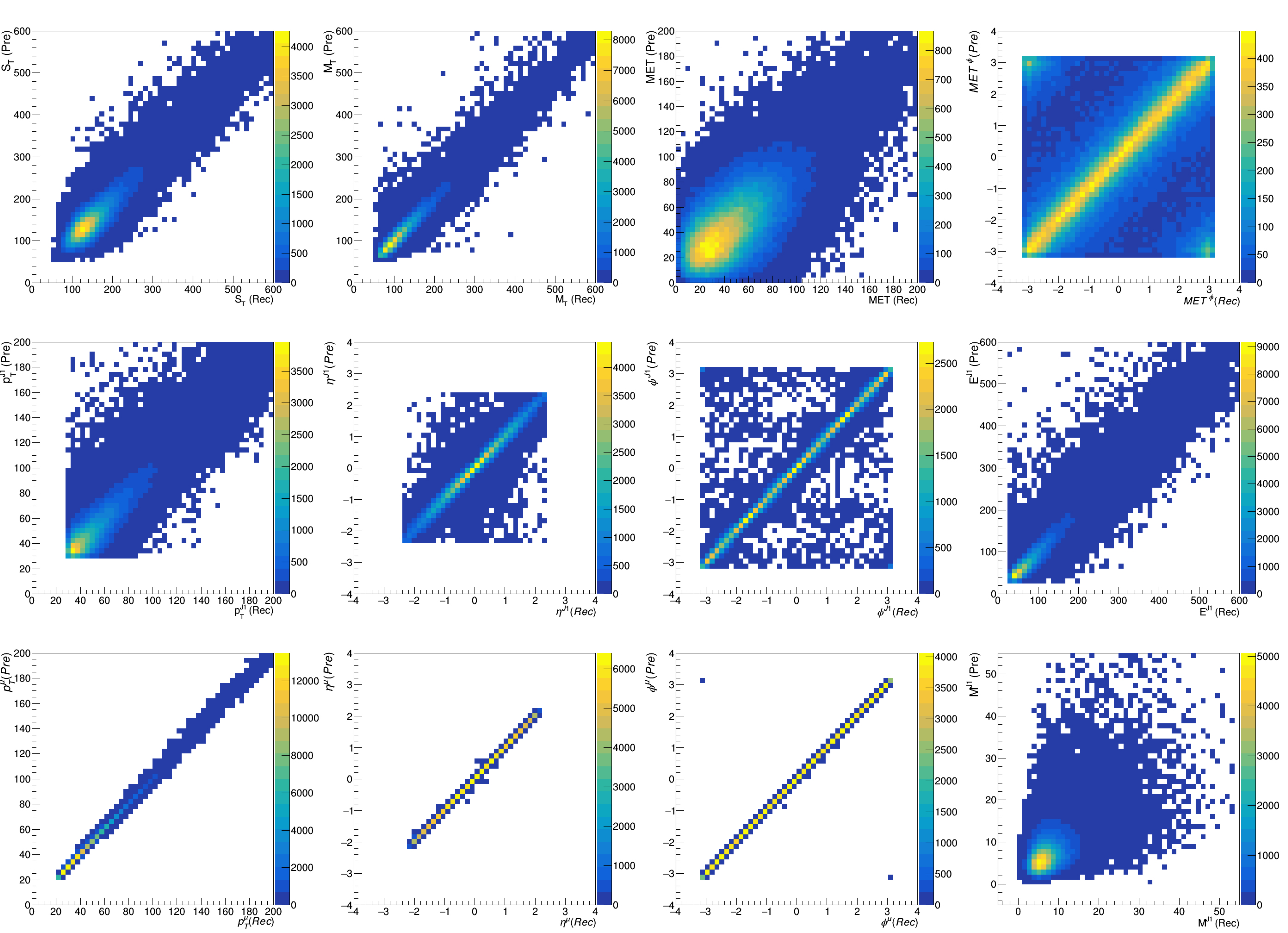}
    \caption{Distribution of auxiliary features predicted by the model as a function of the corresponding quantities from detector simulation.\label{fig:2D_derived}}
\end{figure}

Figures~\ref{fig:2D_input}~and~\ref{fig:2D_derived} show the distribution predicted by the model as a function of the corresponding quantities from detector simulation, respectively for input and auxiliary features. Both the reconstructions techniques start from the generator-level information and model the detector response through a set of random degrees of freedom. The strong correlation and the symmetric distribution around the diagonal demonstrate that, to a large extent, the two event representations are equivalent. 
 
\section{Scaling with dataset size}
\label{app:five_time_stat}

\begin{figure}
    \centering
    \includegraphics[width=0.8\textwidth]{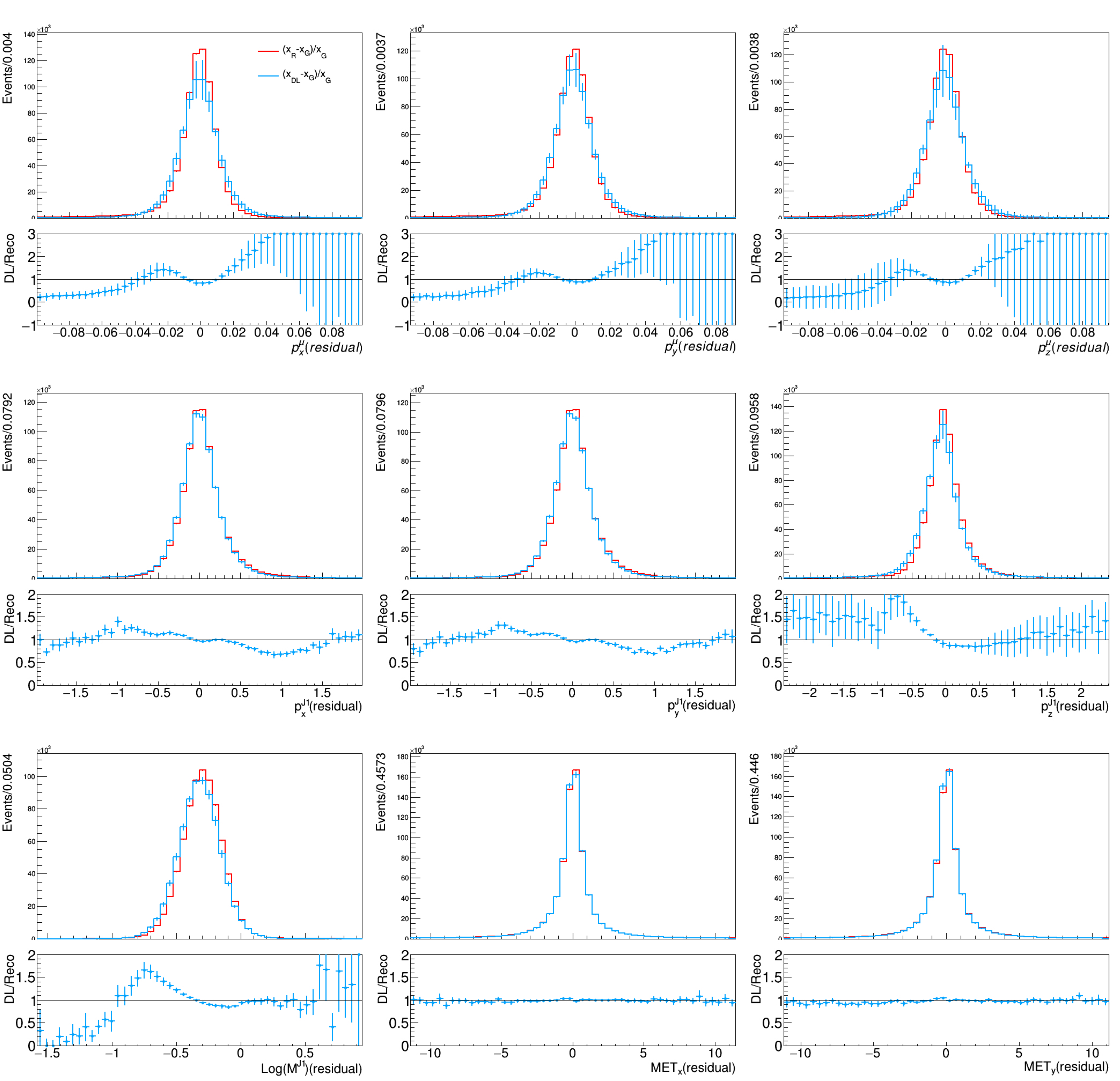}
    \caption{Predict on an inference dataset five times larger than the training dataset. Relative residual distribution for reconstructed and model-predicted quantities in the feature vector, computing with respect to the reference input.
    \label{fig:residual_input_more}}
\end{figure}

\begin{figure}
    \centering
    \includegraphics[width=\textwidth]{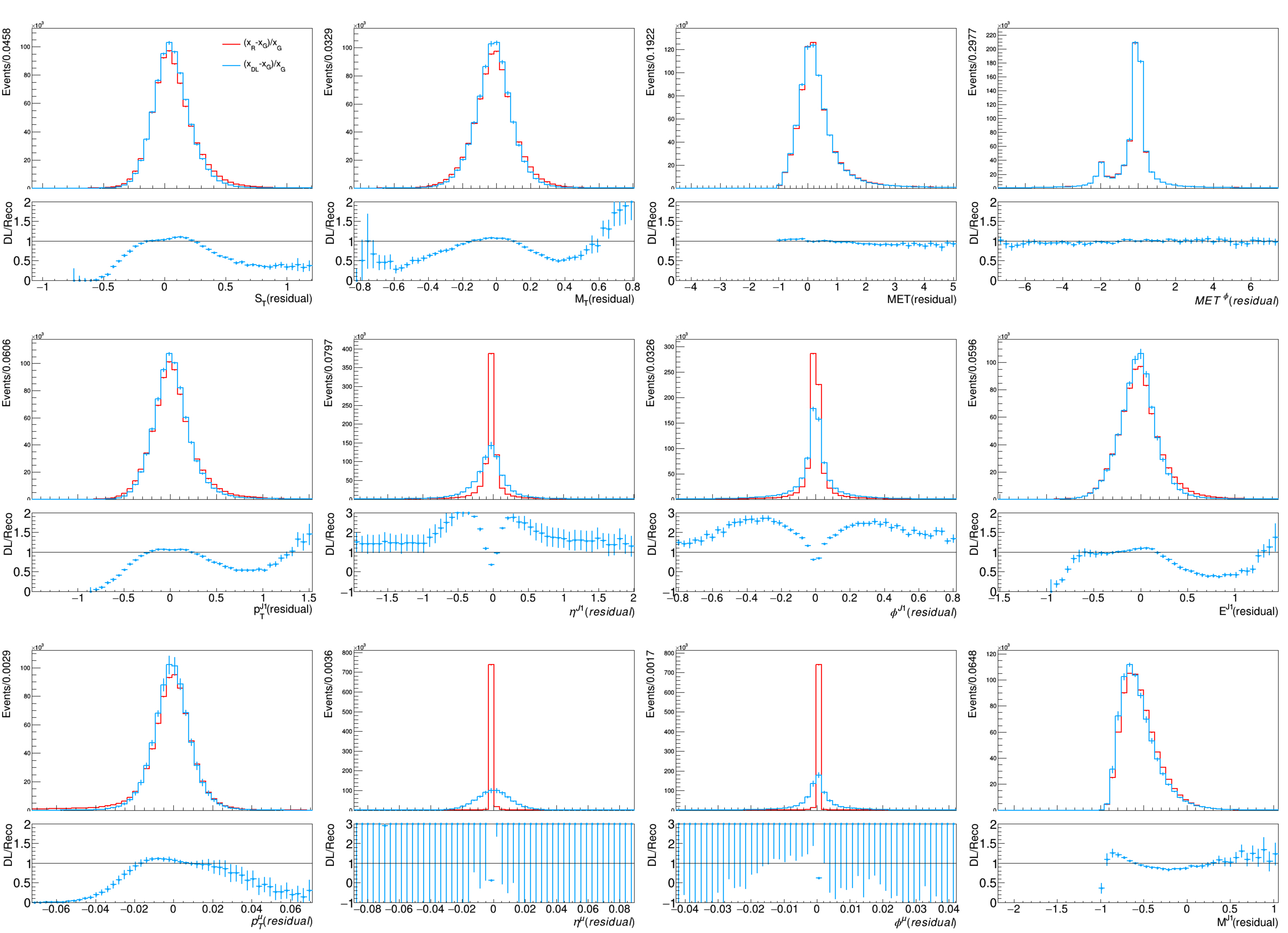}
    \caption{Predict on an inference dataset five times larger than the training dataset. Relative residual distribution for reconstructed and model-predicted auxiliary quantities, computing with respect to the reference input. 
    \label{fig:residual_derived_more}}
\end{figure}

Figures~\ref{fig:residual_input_more}~and~\ref{fig:residual_derived_more} show the comparison between reconstructed and generated quantities with five times more data, computed from detector simulation and processing the generator-level event with our model. Qualitatively, these distributions agree with those of Figs.~\ref{fig:residual_input}~and~\ref{fig:residual_derived}, i.e., no accuracy deterioration is observed due to the scaling of the dataset size. This fact proves the robustness of the proposed methodology and its effectiveness for data augmentation.

\bibliographystyle{mine}
\bibliography{biblio}

\end{document}